# Categorizing h-index variants

M. Schreiber[*], C.C. Malesios[#], and S. Psarakis[#]

[*] *Institut für Physik, Technische Universität Chemnitz, 09107 Chemnitz, Germany*
[#] *Department of Statistics, Athens University of Economics and Business, 76 Patission St, 10434 Athens, Greece*

Utilizing the Hirsch index $h$ and some of its variants for an exploratory factor analysis we discuss whether one of the most important Hirsch-type indices, namely the g-index comprises information about not only the size of the productive core but also the impact of the papers in the core. We also study the effect of logarithmic and square-root transformation of the data utilized in the factor analysis. To demonstrate our approach we use a real data example analysing the citation records of 26 physicists compiled from the Web of Science.

## 1. Introduction

Prior to the introduction of the h-index by *Hirsch (2005)*, individual scientific performance was assessed using traditional bibliometric indicators, such as the number $N$ of articles published or the number $S$ of all citations received by all the published articles. Many other indicators have been devised, such as the average number of citations per article, percentage of highly cited articles, impact indicators based on the impact factor of publication journals of the researcher etc. (For more on traditional bibliometric indicators see e.g. *Costas and Bordons, 2007; Van Leeuwen, Visser, Moed, Nederhof, Van Raan, 2003*). The h-index, based on the set of most cited articles published by a researcher and the citations received by those articles has immediately attracted the attention of the scientific community for assessing the scientific performance of a researcher based on bibliometric data. It is defined by: "A scientist has index *h* if *h* of his *N* papers have at least *h* citations each, and the other (*N* - *h*) papers have at most *h* citations each".

Besides its popularity, a lot of criticism has been raised (*see, e.g., Adler, Ewing, & Taylor, 2009; Schreiber, 2007; Vinkler, 2007; Meho, 2007*), and various modifications and generalizations have appeared (*see, e.g., Egghe, 2006a; Jin, Liang, Rousseau, & Egghe, 2007; Sidiropoulos, Katsaros, & Manolopoulos 2006; Tol, 2009*). There are a number of situations in which the h-index may provide misleading information about a scientist's output. For instance the lack of sensitivity of *h* to highly-cited papers in the h-core (the *h* most cited papers that are counted for *h* because they received *h* or more than *h* citations) is a frequently noticed disadvantage. To relax this "robustness" of the h-index, among others the g-index (*Egghe, 2006b*), the A-index (*Jin, 2006*), the R-index (*Jin et al., 2007*) and the $h_w$-index (*Egghe & Rousseau, 2008*) were proposed. For a comprehensive and critical review of the h-index and similar indices see *Panaretos & Malesios (2009)* and *Schreiber (2010b)*. In the same vein, *Alonso, Cabrerizo, Herrera-Viedma and Herrera (2009)* describe a large set of h-type indices, along with a presentation of the attempts for the standardisation of h-index for measuring scientific performance between different fields of research.

After a comparison of some of the more important variants by means of a factor analysis (FA), *Bornmann, Mutz, & Daniel (2008)* came to the conclusion that there are two types of indices, one type of indices that "describe the most productive core of the output of a scientist and tell us the number of papers in the core" (p. 836) while "the other indices depict the impact of the papers in the core" (p. 836). The authors propose to use for evaluation purposes any pair of indices belonging to these two distinct groups, because each one of the two selected indices should represent one of the two dimensions. In particular, the h-index **and** the g-index were classified as belonging to the first category, while *A* and *R* certainly belong to the second group. To make the data suitable for the FA, the authors applied a logarithmic transformation to the raw data which moreover were shifted by 1 to avoid zero values. Along the same lines, *Costas & Bordons (2008)* performed exploratory factor analysis (EFA) on a dataset of 10 bibliometric indices, including *h* and *g* and also reported a factor structure where both indices, *h* and *g*, load on the same factor solely. Here too, however, the authors have applied a transformation to the data prior to the FA, namely the square root transformation.

Although the aforementioned results are based on elaborate FA, at a closer look the distinction is not so convincing. *Schreiber (2010a)* has argued that the g-index also measures the impact of the papers in its core in the same way as the A-index does for the h-core and, therefore, *g* should be included in the second group as well; but, this does not mean that it should be deleted from the first group. This would make the g-index rather unique among the various variants of the h-index and could be an explanation why "*g* has received most attention, whereas many other derivatives of the h-index have had little response" (*Bornmann & Daniel, 2009c, p. 5*). We investigate this in the following by an analysis of the citation records of 26 physicists, which were previously studied (*Schreiber, 2008 and 2010b*). Specifically, we will use EFA to discuss our assumption that the g-index can be classified as a bibliometric index that measures both the "quantity of the productive core" and the "impact of the productive core".

**2. Data**

Data for the subsequent analysis were compiled between January and February 2007 from the Science Citation Index provided by Thomson Scientific in the Web of Science (WoS). As specified in *Schreiber (2007)*, the 26 datasets comprise the citation records of all full, associate, and assistant professors from the Institute of Physics at Chemnitz university including recently retired professors. The datasets are labeled A, B, C, ….., Z in conformity with the previous analysis (*Schreiber, 2007*). The same data were utilized for an investigation of the g-index in comparison with the h-index, the A-index, and the R-index (*Schreiber, 2008*).

Details of the determination of the datasets have been described elsewhere (*Schreiber, 2007*) in particular with respect to the precision problem, i.e. to establish that the considered publications have really been (co)authored by the investigated scientists and not by colleagues with the same name and the same initials. One may argue that the current analysis is based on a small sample from a single university. On the other hand, however, the great care that has been given by the author in establishing a correct database, that includes scientific staff of a typical institution – in contrast to most of studies that analyse the publication records of very prominent persons – constitutes a representative sample from an average institute and thus we believe that this sample has its specific merit. In the current article we utilize 7 Hirsch-type indices (see Appendix A for the actual values and Appendix B for a brief definition of the 7 indices) which were also used by *Bornmann et al. (2008),* namely $h$, $m$, $g$, $h(2)$, $A$, $R$, $h_w$ thus enabling us to compare between the results of the current analysis and the one conducted by *Bornmann et al. (2008).* The selected indices do not involve any alteration of the original citation data like taking the age of the paper or the productive age of the scientist into account, such as for the AR-index or the m-quotient, or considering the number of coauthors as it is done for $h_m$.

## 3. Methodology - Overview

The goal of EFA is to identify the latent structure present in a set of observed variables, called the factors or latent variables, which are not directly measurable but represent certain features inherent in the data. In this way, EFA reduces the dimensionality of the data to a few representative factors, and therefore summarizes the multivariate information in a simpler form.

There is a steadily increasing literature on applications of EFA in scientometrics in recent years. For example, in a comparative study of some of the most important h-type indices proposed in the literature, *Bornmann et al. (2008)* perform an EFA using as observed variables the g-, h(2)-, A-, R-, AR-, $h_w$-, m-indices and m-quotient in addition to the h-index to identify possible subsets among these indicators that are more correlated to each other. The data described the 414 scientists from biomedicine who applied for BIF fellowships between years 1990 and 1995. The study concludes that more than 95% of the variability in the data is explained by two factors. The first factor was recognized to describe the size of the core (represented by *h*, *g*, *h*(2) and the m-quotient), whereas the second factor was recognized to describe the impact of the papers in the core (reflected by *A*, *R*, *AR*, $h_w$, and the m-index).

In a follow-up study, *Bornmann, Mutz, & Daniel (2009a)* re-ran the aforementioned EFA, adding to the previously described h-type indicators the two standard indicators in scientometrics, i.e., the total number *N* of articles published and the sum *S* of all citations received by these articles. Again the same factors were identified. As a tool for assessing the research performance of scientists, the authors proposed the use of any pair of indicators from the two factors, i.e. one indicator that is related to the number of papers in the researcher's productive core and one indicator that is related to the impact of the papers in the researcher's productive core.

Similar results were reported in *Bornmann, Mutz, Daniel, Wallon, & Ledin (2009b)* as a result of an EFA with, additionally to the previous analyses, two further variants recently proposed, namely the $h_m$-index (*Schreiber, 2009a*) and the Maxprod-index (*Kosmulski, 2007*). The $h_m$-index was found to relate to the number of papers in the productive core (first factor), Maxprod to the impact of the papers in that core (second factor). The data used for this analysis consisted again of young researchers, namely 693 applicants to the Long-Term Fellowship (LTF) programme of the European Molecular Biology Organization.

*Costas & Bordons (2007)* also implemented EFA to investigate possible associations of the h-index with other measures of scientific research such as *N*, *S*, average number of citations per article *C = S/N*, percentage of highly cited papers, the median impact factor, the normalized position of the publication journal, the relative citation rate (RCR) and the percentage of papers with an RCR above 1. Specifically, the authors used WoS data on the publication output of 348 Spanish scientists in the field of natural resources between 1994 and 2004. Here, four factors extracted from the FA explained 93% of the total variance in the data. The first factor (explaining 29% of total variability) comprised *N* and *S* in addition to *h*, while the remaining three factors consisted of relative indicators of quality and quantity. Since the number of publications and citations were characterized as absolute indicators of quantity and impact, respectively, according to the authors' opinion the h-index is confined to explain only a small portion of the information about a researcher's work, leaving unexplained other important aspects of scientific performance, conveyed by the other relative indicators included in the analysis.

The subsequent results of *Costas and Bordons (2008)* indicated that both, *h* and *g* express the same dimension of the research output of a scientist, together with *N* and *S*. In a complementary analysis, however, the authors deduce that the g-index is more sensitive in assessing the performance of selective scientists (i.e. scientists with low production of papers but of high impact) in comparison to the h-index tending to favor the big producers.

In another recent application of data reduction methodology to bibliometric measures, *Hendrix (2008)* analyzed bibliometric data obtained from the WoS on the faculty of the Association of the American Medical Colleges member schools, covering the period between 1997 and 2007, and a total of 123 researchers. The bibliometric variables collected, such as *N*, *S*, *C*, average number of citations per faculty member, combined with other institutional information (e.g., faculty size, total funding), and also an h-type index [the impact index of *Molinari & Molinari (2008)*], were analyzed using the method of principal components analysis (PCA). Three factors were extracted from the fit of the PCA model, the first of which accounted for the 31% of total variance, and was found to be highly correlated with the size-dependent measures (i.e. *N* and *S*), whereas the second factor which explained 30% of total variance is associated with *C*, the impact index, and percentage of articles with zero citations. The third factor essentially accounted for the size-independent bibliometric variables, such as the average number of articles per faculty member and average number of citations per faculty member.

In this paper we employ an EFA model in order to derive categorizations of the h-index and some of its variants. With 26 datasets the sample size is relatively small for the FA. But when certain conditions are fulfilled, in particular when communalities are high and the number of factors is small, then reliable FA results can be obtained even with very small sample sizes like 10 datasets as *Preacher & MacCallum, 2002* have shown (see also *MacCallum, Widaman, Zhang and Hong, 1999*). The importance of large communalities in diminishing the effects of small datasets for performing FA has been already pointed out in a series of studies (e.g. *Pennell, 1968; Velicer, Peacock and Jackson, 1982; Velicer and Fava, 1998*). *MacCallum, Widaman, Preacher and Hong (2001)* added to the previous analyses by showing that communality level is still the most dominant factor in the stability of FA results, even under the most realistic situations. Specifically *MacCallum et al. (1999)* (p. 96) suggested that all communalities should be greater than 0.6, or the mean level of communality should be at least 0.7. As we utilize only two factors and as the determined communalities are extremely high (much higher than 0.9 for most variables, and a mean communality level as high as 0.978 for the raw data), we are confident that the subsequent analysis produces valid results.

## 4. Distribution of the Index Values

As mentioned above, in our EFA model the indices used as input for the analysis were most of the indices used for the FA by *Bornmann et al. (2008)*. AR, $h_m$, and the m-quotient were not included, because they require an alteration of the original citation data.

Additionally to the raw index values $x$ the logarithmically transformed data $\ln(x)$, and the logarithmically transformed shifted data $\ln(x+1)$ were utilized. *Bornmann et al. (2008)* used logarithmized shifted data, and thus it is of interest to check if there are any discrepancies in the results between the raw indices and the transformed ones[1]. We also analyzed the square-root transformed data $\sqrt{x}$ in analogy to *Costas & Bordons (2008)*.

*Bornmann et al. (2008)* and *Costas & Bordons (2008)* have utilized transformations to make the data more suitable for the FA, since EFA techniques require that the variables should be normally distributed. In the case of our datasets there is no need, however, for applying such transformation, since the non-parametric Kolmogorov-Smirnov test for normality has shown that none of the 7 items deviate from normality at a 5% level of statistical significance (see Table 1 for the results of the test). This corresponds to the previous observation (*Schreiber, 2009b*) that $h$ and 3 of its variants are approximately normally distributed for the present 26 datasets.

---

[1] We have also performed the EFA of the 7 bibliometric indices, where instead of the h- and g-indices we use the interpolated $\tilde{h}$-index and the interpolated $\tilde{g}$-index. We hypothesized that the changes of the results by using the interpolated indices will not be significant. Indeed, by comparing the results of loadings of both EFA models we saw that the loadings are almost identical for both analyses. Slight differences were observed only in the EFA using the matrix of the $\ln(x)$.

**Table 1:** One-sample Kolmogorov-Smirnov test of the raw indices

|                           |   h   |   m   |   g   | h(2)  |   A   |   R   |  $h_w$  |
|---------------------------|-------|-------|-------|-------|-------|-------|---------|
| Mean                      | 14.88 | 25.58 | 23.96 |   5   | 33.55 | 22.18 | 19.04   |
| Median                    |  14   | 23.25 |  22   |   5   | 29.5  | 20.2  | 17.75   |
| Standard deviation        | 6.92  | 12.95 | 11.99 |  1.6  | 17.80 | 10.82 |  9.20   |
| Kolmogorov-Smirnov D (*)  | 0.186 | 0.198 | 0.202 | 0.230 | 0.217 | 0.199 | 0.186   |
| p-value (*)               | 0.332 | 0.260 | 0.241 | 0.125 | 0.174 | 0.255 | 0.331   |
| Kolmogorov-Smirnov D (**) | 0.100 | 0.114 | 0.094 | 0.189 | 0.096 | 0.090 | 0.092   |
| p-value (**)              | 0.955 | 0.887 | 0.976 | 0.312 | 0.970 | 0.983 | 0.980   |

*(*) Test distribution is normal.*
*(**)Test distribution is Student.*

Closeness of mean and median values of all indices as shown in Table 1 is also indicative of the approximate normality of the distributions. Nevertheless, transformed data were also investigated. In this case the mean and median values are nearly identical and the p-values of the Kolmogorov-Smirnov test are extremely large, see Tables A2 and A3 in the Appendix. We conclude that for the transformed data the assumption of normality is fulfilled with very high confidence. However, as Table 1 shows that we can be confident that already the non-transformed index values are approximately normally distributed, we proceed with our investigation of the non-transformed data. We also studied the transformed data in the subsequent analysis to demonstrate that they should be used with caution. The EFA of the raw data will provide us with a baseline for assessing the effects of the three types of transformations.

Due to the small number of datasets one would expect that the index values are better described by Student's *t*-distribution. We have performed the respective Kolmogorov-Smirnov test and the results in Tables 1, A2, and A3 confirm that especially the untransformed data are even better described by the *t*-distribution than by the normal distribution. For the transformed data the same conclusion can be drawn.

One possible reason for which – in contrast to the data of *Bornmann et al. (2008)* – our data of each of the 7 indices are approximately normally distributed is the diversity of the status of the selected researchers. Indeed, among the 26 researchers there are young researchers (with comparatively low index scores) as well as senior professors with high values of most of their indices. On the other hand, *Bornmann et al. (2008)* study the data of young researchers, whose index values (especially the values of *h* and *h*(2)) are small and are concentrated within a very narrow field of values, with the direct consequence of giving extremely skewed distributions. It might also be an issue that the discrimination problem is more severe, because due to the small values there occur very many ties for these indices which are restricted to integer values. In our current analysis the discrimination problem also occurs, in particular for *h*(2) where only 7 different values result for the 26 datasets. This leads to problems when one tries to compare with the normal or the *t*-distribution. It is therefore not surprising that the respective *p*-values in Tables 1, A2, and A3 are relatively small. However, if one applies a piecewise linear interpolation of the rank-frequency function one can interpolate between the integer values and thus discriminate all 26 datasets (*Schreiber, 2008*). For these interpolated values of *h*(2) the resulting *p* = 0.682 (test with normal distribution) and *p* = 0.997 (test with t-distribution) are significantly larger than *p* = 0.125 and 0.312 for the non-interpolated *h*(2)-index. Similar observations have been made for the transformed data.

The differences in the two analyses are obvious considering that the 414 scientists investigated by *Bornmann et al. (2008)* have altogether published a total number of 1,586 papers, receiving a total number of 60,882 citations which means on average 3.8 papers and 147 citations per scientist, whereas the 26 scientists in our study have published a much higher number of papers (2,373 papers) that received 25,554 citations, i.e. on average 91 papers and 983 citations per person. The median values of 2, 2, and 3 for *h*(2), *h*, and *g*, respectively, show that the discrimination problem is quite severe, because at least half of the 414 scientists share only 3, 3, or 4 values of the respective indices. The more recent study by *Bornmann et al.* (*2009b*) comprised 693 scientists with 3,351 papers, i.e. 4.8 papers per scientist which received 219 citations per scientist on average. In this case, the median values of 3, 4, and 4 for *h*(2), *h*, and *g* again point to a substantial discrimination problem.

## 5. Exploratory Factor Analysis

For their EFA *Bornmann et al. (2008, 2009b)* used a maximum-likelihood factor extraction procedure which usually gives better and more robust estimates than the standard least-squares ansatz. However, for small sample sizes it has been argued that the least-squares method performs better (*Ihara & Okamoto, 1985*). For the FA we thus utilized least squares and a rotated varimax transformation to make our results more easily interpretable (statistical package SPSS 15.0 was utilized for the analysis; *SPSS, 1999*). In order to confirm the suitability of implementing EFA for the specific data and items selected, the Kaiser-Meyer-Olkin (KMO) measure of model adequacy was used (*Kaiser, 1974)*, which gave the acceptable value of 0.737 (see Table 2). The factor loading matrices for the factor models with the 7 indices can be found in Table 3. The results of the analysis indicate the existence of two factors as the best solution for explaining the variability in the data, as in *Bornmann et al. (2008)*.

**Table 2:** KMO test

|         | Raw indices $x$ | $\ln(x)$ | $\ln(x+1)$ | $\sqrt{x}$ |
|---------|-----------------|----------|------------|------------|
| KMO     | 0.737           | 0.830    | 0.813      | 0.744      |
| p-value | < 0.001         | < 0.001  | < 0.001    | < 0.001    |

**Table 3:** Varimax rotated loading matrices (applying Kaiser normalization) for the 4 EFA models with values above 0.6 given in bold face

| Indices | Raw indices $x$ | | $\ln(x)$ | | $\ln(x+1)$ | | $\sqrt{x}$ | |
|---------|------|------|------|------|------|------|------|------|
|         | Component 1 | Component 2 | Component 1 | Component 2 | Component 1 | Component 2 | Component 1 | Component 2 |
| $h$     | **0.842** | 0.522 | **0.825** | 0.496 | **0.828** | 0.496 | **0.841** | 0.504 |
| $m$     | **0.752** | 0.597 | **0.721** | 0.525 | **0.728** | 0.524 | **0.742** | 0.561 |
| $g$     | **0.722** | **0.691** | **0.705** | **0.705** | **0.707** | **0.702** | **0.717** | **0.694** |
| $h(2)$  | **0.789** | 0.572 | **0.843** | 0.514 | **0.839** | 0.520 | **0.812** | 0.548 |
| $A$     | 0.536 | **0.844** | 0.491 | **0.871** | 0.494 | **0.870** | 0.514 | **0.858** |
| $R$     | **0.718** | **0.695** | **0.708** | **0.703** | **0.709** | **0.701** | **0.716** | **0.696** |
| $h_w$   | **0.732** | **0.681** | **0.719** | **0.694** | **0.722** | **0.691** | **0.728** | **0.685** |
| Eigenvalues | 3.755 | 3.094 | 3.667 | 3.017 | 3.688 | 3.01 | 3.739 | 3.042 |

Table A4 shows the variance (communality) of each item explained by the factors extracted from the 4 models. Both factors accounted for 97.83%, 95.48%, 95.68%, and 96.87% of the total variance in the raw data, the log-transformed data, the log-transformed shifted data, and the square-root transformed data, respectively. For the raw data, the first factor explained 53.64% of the total variance present in the data, whereas the second factor explained 44.19%. We have used the categorization of the indices among the factors by using factor loadings greater than 0.6 in agreement with *Bornmann et al.* (*2008*). As Table 3 shows, for the raw data the indices loading on the first factor are $h$, $m$, $h(2)$, $h_w$, $g$ and $R$. The factors loading on the second factor are $g$, $A$, $R$, and $h_w$. We cannot assign – in the way *Bornmann et al. (2008)* did – indices loading on the first factor as indicating the number of papers in the productive core of the researchers' outputs, because $m$, $h_w$, and $R$ are surely based on the number of citations in the core. On the other hand, all the indices loading on the second factor reflect the impact of the papers in that core. We observe, that $g$, $R$, and $h_w$ are included in both factors. Indices solely loading on the first factor are $h$, $m$ and $h(2)$, whereas the only index loading solely on the second factor is $A$. In principle one could force a distinction by fine-tuning the threshold: a value 0.7 would lead to a separation of all indices except *A* loading on the first factor and only *A* on the second. But this is not convincing, because the factor loadings for the two components are rather close for *g*, *R,* and $h_w$.

By utilizing the transformed data we obtain rather similar results as Table 3 shows. The results of the FA utilizing the square-root transformation are in accordance with the factor loadings for the raw data. For the log-transformed data ln(*g*), ln(*R)*, and ln($h_w$) the loadings on both factors are even closer than for the raw data, confirming the conclusion that these indices load on both factors. Shifting the data by 1 before logarithmizing yields only minor differences in comparison to the unshifted logarithmized data. Thus our categorization differs from that obtained by *Bornmann et al.* (*2008*)*,* because there a clear separation into two groups appears.

**Table 4:** Promax oblique rotated loading matrices for the 4 EFA models with values above 0.6 given in bold face

| Indices | Raw indices *x* | | ln(*x*) | | ln(*x*+1) | | $\sqrt{x}$ | |
|---|---|---|---|---|---|---|---|---|
| | Component | | Component | | Component | | Component | |
| | 1 | 2 | 1 | 2 | 1 | 2 | 1 | 2 |
| *h* | **0.842** | 0.183 | **0.824** | 0.173 | **0.829** | 0.170 | **0.848** | 0.164 |
| *m* | **0.663** | 0.350 | **0.661** | 0.279 | **0.671** | 0.272 | **0.670** | 0.311 |
| *g* | 0.556 | 0.504 | 0.519 | 0.543 | 0.524 | 0.537 | 0.546 | 0.515 |
| *h*(2) | **0.732** | 0.290 | **0.838** | 0.186 | **0.829** | 0.196 | **0.778** | 0.246 |
| *A* | 0.187 | **0.848** | 0.110 | **0.914** | 0.114 | **0.910** | 0.148 | **0.882** |
| *R* | 0.547 | 0.514 | 0.524 | 0.538 | 0.528 | 0.534 | 0.542 | 0.520 |
| $h_w$ | 0.577 | 0.484 | 0.546 | 0.518 | 0.552 | 0.512 | 0.568 | 0.495 |
| Eigenvalues | 6.163 | 5.783 | 5.918 | 5.565 | 5.945 | 5.574 | 6.068 | 5.672 |

The above utilized varimax rotation method is an orthogonal rotation which assumes that the factors in the analysis are uncorrelated. We have also applied an oblique rotation method, which allows factors to be correlated. Such techniques have been favored in comparison with orthogonal rotations (see e.g. *McCroskey & Young, 1979*). Specifically we employed the promax rotation with exponent 3. *Tataryn, Wood & Gorsuch (1999)* advise exponents 2, 3, or 4 for optimal results. In our case we have also tested exponents 2 and 4 and did not get large differences. Results are presented in Table 4. Again *h*, *m*, *h*(2) load on the first factor and *A* on the second factor. Once more, *g*, *R*, $h_w$ load on both factors, although now their loadings are remarkably smaller than in Table 3 and turn out to be below the threshold value 0.6. The same observations can be made for the transformed data. In conclusion, the oblique rotation method does not improve the categorization.

## 6. Expanded Data Base

As mentioned above, in subsequent FAs comprising also $N$ and $S$ Bornmann et al. (2009a; 2009b) again found a distinct categorization with respect to the two factors, with $h$, $m$-quotient, $g$, $h(2)$, $h_m$, and $N$ loading on the quantity dimension, and $A$, $m$, $R$, $AR$, $h_w$, Maxprod and $S$ loading on the impact dimension. *Bornmann et al. (2009b)* conclude that "for measuring quantity and quality of research performance, the h-index and its variants do not necessarily have to be used". However, the authors also note that their findings are referring to a dataset from a specific field of research, i.e. the area of biomedicine, and that further analyses are required before generalizing these conclusions. This has prompted us to re-run our EFA including $N$ and $S$. The results are presented in Table 5. (The KMO test yields acceptable values between 0.799 for the raw indices and 0.844 for the log-transformed data. All p-values are below 0.001). Surprisingly, now all indices load on one component and $N$ on the other. $S$ also loads on the first factor but the value of the second factor loading is close to the threshold. This means that for our datasets $N$ and $S$ cannot be used in the EFA for distinguishing different categories of the index variants. It is interesting to note that the transformations yield changes in the loading matrix, for all the transformed data $h$ clearly loads on both factors and $S$ loads evenly on both factors.

**Table 5:** Varimax rotated loading matrices as in Table 3, but comprising also $N$ and $S$

| Indices | Raw indices $x$ | | $\ln(x)$ | | $\ln(x+1)$ | | $\sqrt{x}$ | |
|---|---|---|---|---|---|---|---|---|
| | Component | | Component | | Component | | Component | |
| | 1 | 2 | 1 | 2 | 1 | 2 | 1 | 2 |
| $h$ | **0.815** | 0.545 | **0.672** | **0.711** | **0.680** | **0.705** | **0.738** | **0.640** |
| $m$ | **0.855** | 0.436 | **0.766** | 0.440 | **0.770** | 0.441 | **0.795** | 0.476 |
| $g$ | **0.902** | 0.434 | **0.847** | 0.528 | **0.849** | 0.525 | **0.869** | 0.494 |
| $h(2)$ | **0.846** | 0.463 | **0.779** | 0.566 | **0.784** | 0.563 | **0.805** | 0.535 |
| $A$ | **0.919** | 0.301 | **0.914** | 0.320 | **0.914** | 0.319 | **0.914** | 0.321 |
| $R$ | **0.907** | 0.422 | **0.853** | 0.520 | **0.855** | 0.516 | **0.876** | 0.481 |
| $h_w$ | **0.898** | 0.443 | **0.854** | 0.522 | **0.855** | 0.520 | **0.868** | 0.498 |
| $N$ | 0.375 | **0.926** | 0.348 | **0.890** | 0.349 | **0.891** | 0.363 | **0.889** |
| $S$ | **0.765** | 0.592 | **0.702** | **0.712** | **0.703** | **0.710** | **0.715** | **0.687** |
| Eigenvalues | 6.123 | 2.561 | 5.269 | 3.242 | 5.306 | 3.220 | 5.579 | 3.013 |

In order to get a clearer distinction, we performed a further EFA including again two standard bibliometric indicators, namely the total number $N$ of papers and the average number $C$ of citations to all publications, in addition to the 7 indices of our first analysis. We now use $C$ rather than the sum $S$ of all citations, because $S$ can be expected to correlate stronger with the quantity of publications than $C$. ("More papers attract more citations.") The results of the analysis are presented in Tables 6 and A5. (The KMO test yields acceptable values between 0.758 for the raw indices and 0.819 for the log-transformed data. All p-values are below 0.001).

The high loadings of $N$ and $C$ on the first and the second factor, respectively, mean that by including $N$ and $C$ into the analysis we have successfully enforced a distinction of the quantity and the quality dimension. However, unfortunately this does not lead to a clear discrimination of the investigated indices, since the factor loadings are far from the ideal case where each item has a relatively large loading on one factor and a near-zero loading on the other factor. Rather most indices appear in both dimensions if we use the threshold of 0.6 for the categorization of factor loadings. This clearly indicates strong differences in comparison to the results derived by *Bornmann et al. (2008, 2009a, 2009b)*. For most indices the loadings on the first factor are larger than those on the second factor so that increasing the threshold to the value 0.7 would distinguish $A$ and $C$ loading on the second factor and all others on the first factor for the raw data. The results for the square-root transformed data are somewhat different, with $\sqrt{m}$, $\sqrt{g}$, $\sqrt{R}$, and $\sqrt{h_w}$ clearly loading on both factors, and even $\sqrt{h(2)}$ loading on both factors. More conspicuous are the deviations for the log-transformed data, because now $\ln(m)$, $\ln(g)$, $\ln(R)$, and $\ln(h_w)$ are loading even more strongly on the second factor.

**Table 6:** Varimax rotated loading matrices as in Table 3, but comprising also $N$ and $C$

|  | Raw indices $x$ | | $\ln(x)$ | | $\ln(x+1)$ | | $\sqrt{x}$ | |
|---|---|---|---|---|---|---|---|---|
| Indices | Component | | Component | | Component | | Component | |
|  | 1 | 2 | 1 | 2 | 1 | 2 | 1 | 2 |
| $h$ | **0.827** | 0.536 | **0.812** | 0.537 | **0.810** | 0.542 | **0.808** | 0.552 |
| $m$ | **0.729** | **0.620** | 0.591 | **0.658** | 0.597 | **0.659** | **0.666** | **0.644** |
| $g$ | **0.745** | **0.666** | **0.689** | **0.722** | **0.689** | **0.722** | **0.708** | **0.704** |
| $h(2)$ | **0.768** | 0.591 | **0.722** | **0.640** | **0.723** | **0.642** | **0.740** | **0.625** |
| $A$ | **0.601** | **0.769** | 0.500 | **0.825** | 0.502 | **0.824** | 0.544 | **0.803** |
| $R$ | **0.734** | **0.679** | **0.681** | **0.730** | **0.681** | **0.730** | **0.698** | **0.715** |
| $h_w$ | **0.753** | **0.658** | **0.688** | **0.725** | **0.689** | **0.725** | **0.714** | **0.701** |
| $N$ | **0.909** | 0.095 | **0.970** | 0.131 | **0.969** | 0.130 | **0.948** | 0.116 |
| $C$ | 0.162 | **0.986** | 0.124 | **0.992** | 0.124 | **0.992** | 0.138 | **0.991** |
| Eigenvalues | 4.680 | 3.931 | 4.148 | 4.395 | 4.156 | 4.399 | 4.359 | 4.248 |

We have also performed a further EFA comprising the 7 index variants and *N*, *C*, as well as *S*. We do not show the results here, because the loading matrices are very similar to those given in Table 6. This means that again *N* and *C* lead to a clear distinction of the quantity and the quality dimension. The total number *S* of all citations to all publications loads strongly on the same factor as *N* with matrix elements of the order of 0.83 in the 4 EFA models for the varimax and around 0.78 for the promax rotation, i.e. more strongly than all the index variants. We conclude that *S* is not very helpful for the present investigation while *N* and *C* can be used for categorizing the quantity and the quality dimension. Our results are in accordance with the conclusion of *Costas & Bordons (2007, 2008)*, namely that one factor comprises *N*, *S*, and *h* as well as *g*, but not *C*. Our results also agree with the findings of *Hendrix (2008)* who extracted one factor correlated with *N* and *S*, and another factor associated with *C*.

We have also investigated the 7 indices, *N* and *C* with the promax oblique rotation. Results are shown in Table 7. The extremely high loadings of *N* and *C* again show that the distinction of the quality and quantity dimension has been successful. For the raw indices one obtains a clear categorization of *A* and *C* belonging into the second category and all other indices into the first category. These results are in accordance with the varimax rotation method if the higher threshold is applied to the loading matrix in Table 6. We conclude that in this case the oblique rotation leads to a clearer distinction.

Additionally we have followed our EFA with a confirmatory FA (CFA) (*Jöreskog, 1969*) to verify our results from the perspective of statistical testing, since CFA allows for tests of statistical significance for the parameters of the model tested, such as the obtained factor loadings. Tables A6 and A7 present factor loadings and the related tests. By fitting the specific CFA model structure[2] obtained from EFA, we see that all items are significant at the 5% significance level, with only the exception of *N*, which loads somewhat lower in comparison to the EFA results. Generally, however, both tests for factor loadings as well as the respective $R^2$ values indicate the validity of the performed EFA).

---

[2] Estimation for the CFA model was carried out using the LISREL (*Jöreskog and Sörbom, 1999*) software.

**Table 7:** Promax oblique rotated loading matrices as in Table 4, but comprising $N$ and $C$

| Indices | Raw indices $x$ | | ln($x$) | | ln($x$+1) | | $\sqrt{x}$ | |
|---|---|---|---|---|---|---|---|---|
| | Component 1 | 2 | Component 1 | 2 | Component 1 | 2 | Component 1 | 2 |
| $h$ | **0.777** | 0.289 | **0.750** | 0.315 | **0.746** | 0.321 | **0.743** | 0.327 |
| $m$ | **0.623** | 0.433 | 0.439 | 0.547 | 0.445 | 0.545 | 0.534 | 0.497 |
| $g$ | **0.622** | 0.483 | 0.529 | 0.584 | 0.529 | 0.584 | 0.560 | 0.551 |
| $h(2)$ | **0.682** | 0.381 | **0.601** | 0.474 | **0.603** | 0.475 | **0.631** | 0.443 |
| $A$ | 0.403 | **0.670** | 0.264 | **0.782** | 0.266 | **0.780** | 0.322 | **0.738** |
| $R$ | **0.604** | 0.502 | 0.517 | 0.596 | 0.517 | 0.597 | 0.544 | 0.568 |
| $h_w$ | **0.635** | 0.470 | 0.528 | 0.588 | 0.529 | 0.587 | 0.568 | 0.545 |
| $N$ | **1.065** | -0.282 | **1.102** | -0.233 | **1.101** | -0.234 | **1.090** | -0.254 |
| $C$ | -0.224 | **1.126** | -0.252 | **1.133** | -0.252 | **1.133** | -0.243 | **1.132** |
| Eigenvalues | 6.856 | 6.230 | 6.253 | 6.459 | 6.270 | 6.473 | 6.518 | 6.421 |

For the transformed data in Table 7 the same deviations as in Table 6 occur. Although the loadings for $\sqrt{m}$, $\sqrt{R}$, and $\sqrt{h_w}$ are all below the threshold of 0.6, their values for the two components are not so different so that one might conclude that $\sqrt{m}$, $\sqrt{R}$, $\sqrt{h_w}$ load on both factors, if the square-root transformed data are considered. Again for the log-transformed data ln($m$), ln($g$), ln($R$), and ln($h_w$) are loading even more strongly on the second factor. Thus we have again found a remarkable difference between the EFA results for the raw data and for the transformed data. The CFA of the transformed data has also verified the validity of the tested structure, i.e. the structure indicated by the EFA and thus corroborated the deviations, see Tables A6 and A7. This discrepancy is an important finding. Problems when transforming (and especially log-transforming) non-normally distributed data prior to conducting a FA are not new. For instance, *Chapman (1977)* states that "factor analysis is very susceptible to scaling and transformations" and that "comparisons of factor models from the log-transformed and untransformed data showed that after log-transformation widespread factor fusion and factor fission occurred and no factor model was completely satisfactory".

Although transforming the data to demonstrate normality is quite a common procedure, there are other alternatives too, depending on the degree of departure from normality in the specific dataset (see, e.g. *Ferguson & Cox, 1993*). But according to *Muthen & Kaplan (1985)*, some degree of skew and kurtosis in the data is acceptable for conducting EFA, especially when the skew and kurtosis coefficients are below the value of 2. Other alternatives include to retain non-normally distributed items in the analysis if they account to less than 25% of the overall number of items since it is believed that below this percentage the distorted variables will not affect the final solution of the model (*Ferguson & Cox, 1993*). When none of the above holds, based on the results of the current analysis, we recommend using square-root transformations instead of log-transformed datasets.

In the case of bibliometric data, major discrepancies between results of a FA of skewed citation distributions and their log-transformed values (*Leydesdorff & Bensman, 2011*) of journals were found associated with a reduced dimensionality of the final factor solution, classification of the items, and their magnitude as concerns the factor loadings.

Based on these findings, we believe that caution should be exercised when interpreting results based on EFA, if transformations have been applied to the data. In Tables 5-7 we found rather large changes due to the transformations, while there were only minor changes in Tables 3 and 4. Therefore we propose to avoid transformations, unless absolutely necessary or at least to compare the FA results for the raw data and the transformed data.

## 7. Discussion

An important consideration in evaluating research performance of a scientist is the multiple manifestation of his/her work. One of the main disadvantages of the traditional bibliometric indicators, such as the total number of papers or the total number of citations is that they do not account for the quality of scientific research, or that they can be disproportionately affected by a single publication of major influence.

The h-index intends to measure simultaneously quality and quantity of scientific output, taking into account, to some extent, the diversity of scientific research. However, many authors have argued against the use of an index to assess a scientist's work by one single number. *Bornmann et al. (2008; 2009a; 2009b)* propose to utilize two indices, one for the quantity of the productive core of a researcher and one for the impact of the core, basing their arguments on the EFA of specific datasets for young scientists from a specific scientific field of research with certain scientific experience.

With more diverse datasets − comprising scientists of varying scientific age – we got quite different results. We have demonstrated that the logarithmic transformation can cause distortions in the results of EFA and subsequently in the interpretations.

We have shown, that for our datasets, for some of the investigated indices no clear distinction was evident to one of the two dimensions of scientific performance. The argument (*Schreiber, 2010a*) that *g* measures the impact has not been corroborated. The nearly equal factor loadings for *g* in the EFA of the raw data in Table 3 seemed to confirm the assumption (*Schreiber, 2010a*) that the g-index measures both, the quantity and the impact. However, this was not substantiated by the more comprehensive FA in the previous chapter. Our results are in agreement with the findings of *Costas & Bordons (2007, 2008)* and *Hendrix (2008)*. Major differences to previous analyses of *Bornmann et al.* (*2008, 2009a, 2009b*) have been found. They might be due to the small number of papers per scientist in those studies with median values of 2, 3, or 4 for *N, h, g,* and *h*(2) which lead to problems of discrimination between the scientists especially for the integer-valued indices. On the other hand the present study suffers from the relatively small number of only 26 investigated scientists.

Table 3 shows that the 7 indices under consideration cannot be easily categorized into two groups: *A* and *h* appear to define two categories, but most of the other indices load strongly on both factors. The oblique rotation results in Table 4 show the same bevavior, if the lower threshold is applied. Adding standard bibliometric indicators, namely the number of publications *N* and the total number of citations *S*, to the analysis does not help. As shown in Table 5, in this way one can force a distinction into 2 categories, but the second category comprised only *N* and therefore the distinction is not very helpful. Only the consideration of *N* and the average number of citations *C* leads (with or without including *S*) to a more useful separation into two categories. Now *N* defines one group with *h* and *h*(2) while *C* belongs to the other group together with *A*. All other indices load strongly on the first factor, but also show a remarkable share of the second factor, as demonstrated in Table 6. The oblique rotation in Table 7 makes the distinction more obvious.

In conclusion we have clearly identified 2 categories of h-index variants and we propose to use the 2 complementary indices *h* and *A* in the research assessment of an individual scientist. Alternatively, the consideration of *N* and *C* would also reflect these two categories, but due to the usually long tail of the citation statistics these numbers are more difficult to obtain with good precision. Thus *h* and *A* appear to be the better choice. On the other hand, if it is the aim to utilize only a single index, then the variants *m, g, R,* and $h_w$ could be chosen, because they comprise information from both factors, as corroborated in Table 6. For *R* this is not surprising, as *R* combines *h* and *A* by definition. The $h_w$-index is an interesting alternative, but its practical use is hindered by the relatively complicated definition. The m-index is favourable, because of its simplicity, but its calculation needs the precedent determination of *h*, just like $h_w$ and *R* do. In contrast *g* is defined in its own way and in its own right. In this context, it is interesting to note the similarity of the definition of *A* as the average number of citations received by the articles in the h-core and the fact that *g* can be determined by the average number of citations received by the articles included in the g-core. Thus we propose the use of *g*, if the assessment by a single number is wanted.

However, it remains doubtful in principle, whether the scientific achievements of an individual researcher can or should be quantified with one or two numbers. Hirsch (2005) has already expressed the caveat that "a single number can never give more than a rough approximation to an individual's multifaceted profile". Certainly one should better look at more than one indicator, because different citation records can yield the same h-index and/or the same values for some of the variants. This can easily lead to the underestimation of highly-cited but only intermediate-productive scientists (*Costas & Bordons, 2007*). That might influence the publication strategy of those so called selective scientists. There are other problems, because of which using only one or two indicators is inadequate for the research assessment of individuals, like the age effect and the dependence on the research field (*van Leeuwen, 2008*), or the influence of self-citations (*Schreiber, 2009b*). The h-index has also been criticized by a more theoretical consideration, because it may lead to inconsistent results (*Waltman & van Eck, 2009*) at least on the level of analysis of individual scientists. This finding makes the use of the h-index and most or all of its variants questionable from a fundamental point of view.

The current analysis has shown that inferences based solely on specific datasets cannot be and should not be generalized. Other robust techniques are required for unifying and generalizing the results of EFA. More valid inferences could be drawn, for instance, by utilizing bootstrap methods (see *Efron, 1979*) that do not require any distributional assumptions and can perform adequately under both normal and non-normal distributions, even when the sample size is small.

Topics for future research may also include the replication of our analysis for a wider range of h-type indices allowing us to gain from the benefits of a comparative study and the utilization of a more extensive and updated dataset.

## Acknowledgement

We thank Prof. Lutz Bornmann for his suggestion to apply oblique rotation and for further helpful comments.

# Appendix A

**Table A1:** Characteristics of the 26 datasets analyzed in the present study, see also *Schreiber (2010b)*.

| data set | g | h(2) | h | A | m | R | $h_w$ | N | S | C |
|---|---|---|---|---|---|---|---|---|---|---|
| A | 67 | 10 | 39 | 93.9 | 72 | 60.5 | 51.7 | 290 | 5,997 | 20.7 |
| B | 45 | 8 | 27 | 62.6 | 47 | 41.1 | 35.3 | 270 | 3,177 | 11.8 |
| C | 36 | 7 | 23 | 47.3 | 40 | 33.0 | 28.5 | 126 | 1,661 | 13.2 |
| D | 29 | 6 | 20 | 35.5 | 30.5 | 26.6 | 23.6 | 322 | 2,124 | 6.6 |
| E | 37 | 6 | 19 | 62.4 | 38 | 34.4 | 28.2 | 63 | 1,439 | 22.8 |
| F | 26 | 5 | 18 | 32.2 | 29 | 24.1 | 20.7 | 131 | 1,127 | 8.6 |
| G | 23 | 5 | 17 | 28.4 | 26 | 22.0 | 18.3 | 49 | 697 | 14.2 |
| H | 26 | 6 | 16 | 35.9 | 30.5 | 24.0 | 21.4 | 70 | 749 | 10.7 |
| I | 28 | 6 | 15 | 46.1 | 24 | 26.3 | 22.3 | 65 | 885 | 13.6 |
| J | 23 | 5 | 15 | 32.1 | 23 | 21.9 | 18.1 | 51 | 574 | 11.3 |
| K | 21 | 5 | 14 | 27.7 | 26.5 | 19.7 | 16.8 | 79 | 596 | 7.5 |
| L | 22 | 5 | 14 | 30.6 | 23 | 20.7 | 17.8 | 88 | 681 | 7.7 |
| M | 24 | 5 | 14 | 34.0 | 21 | 21.8 | 18.3 | 70 | 726 | 10.4 |
| N | 22 | 5 | 14 | 27.7 | 26 | 19.7 | 17.7 | 72 | 687 | 9.5 |
| O | 19 | 4 | 13 | 22.8 | 18 | 17.2 | 14.9 | 77 | 550 | 7.1 |
| P | 24 | 5 | 13 | 41.5 | 27 | 23.2 | 20.5 | 47 | 631 | 13.4 |
| Q | 15 | 4 | 13 | 17.1 | 17 | 14.9 | 13.0 | 86 | 422 | 4.9 |
| R | 19 | 5 | 12 | 27.0 | 19.5 | 18.0 | 15.4 | 46 | 451 | 9.8 |
| S | 18 | 4 | 12 | 22.8 | 18 | 16.6 | 13.8 | 61 | 439 | 7.2 |
| T | 15 | 4 | 10 | 18.0 | 15.5 | 13.4 | 11.4 | 78 | 375 | 4.8 |
| U | 17 | 4 | 10 | 23.7 | 23.5 | 15.4 | 13.4 | 44 | 351 | 8.0 |
| V | 17 | 4 | 10 | 24.4 | 14.5 | 15.6 | 13.0 | 60 | 389 | 6.5 |
| W | 13 | 3 | 9 | 15.6 | 12 | 11.8 | 10.1 | 53 | 261 | 4.9 |
| X | 18 | 3 | 8 | 35.1 | 10.5 | 16.8 | 14.3 | 35 | 346 | 9.9 |
| Y | 9 | 3 | 7 | 11.0 | 10 | 8.8 | 7.9 | 25 | 116 | 4.6 |
| Z | 10 | 3 | 5 | 17.0 | 23 | 9.2 | 8.5 | 15 | 103 | 6.9 |

**Table A2:** One-sample Kolmogorov-Smirnov test of the log-transformed indices
(Results for the log-transformed shifted indices are very similar.)

| | ln(h) | ln(m) | ln(g) | ln(h(2)) | ln(A) | ln(R) | ln($h_w$) |
|---|---|---|---|---|---|---|---|
| Mean | 2.61 | 3.14 | 3.08 | 1.56 | 3.41 | 3 | 2.86 |
| Median | 2.64 | 3.15 | 3.09 | 1.60 | 3.38 | 3 | 2.88 |
| Standard deviation | 0.42 | 0.45 | 0.43 | 0.30 | 0.47 | 0.43 | 0.42 |
| Kolmogorov-Smirnov D (*) | 0.113 | 0.114 | 0.111 | 0.174 | 0.121 | 0.110 | 0.106 |
| p-value (*) | 0.892 | 0.885 | 0.908 | 0.408 | 0.838 | 0.912 | 0.933 |
| Kolmogorov-Smirnov D (**) | 0.099 | 0.116 | 0.068 | 0.181 | 0.100 | 0.073 | 0.084 |
| p-value (**) | 0.957 | 0.876 | 0.999 | 0.364 | 0.956 | 0.999 | 0.993 |

*(\*) Test distribution is normal.*
*(\*\*)Test distribution is Student.*

**Table A3:** One-sample Kolmogorov-Smirnov test of the square-root transformed data

|  | $\sqrt{h}$ | $\sqrt{m}$ | $\sqrt{g}$ | $\sqrt{h(2)}$ | $\sqrt{A}$ | $\sqrt{R}$ | $\sqrt{h_w}$ |
|---|---|---|---|---|---|---|---|
| Mean | 3.77 | 4.93 | 4.78 | 2.21 | 5.63 | 4.60 | 4.26 |
| Median | 3.74 | 4.82 | 4.69 | 2.24 | 5.43 | 4.49 | 4.21 |
| Standard deviation | 0.82 | 1.15 | 1.10 | 0.34 | 1.39 | 1.04 | 0.95 |
| Kolmogorov-Smirnov D (*) | 0.145 | 0.150 | 0.153 | 0.198 | 0.167 | 0.151 | 0.147 |
| p-value (*) | 0.645 | 0.555 | 0.576 | 0.258 | 0.464 | 0.590 | 0.629 |
| Kolmogorov-Smirnov D (**) | 0.099 | 0.110 | 0.081 | 0.183 | 0.078 | 0.076 | 0.080 |
| p-value (**) | 0.958 | 0.908 | 0.995 | 0.346 | 0.997 | 0.998 | 0.996 |

*(*) Test distribution is normal.*
*(**) Test distribution is Student.*

**Table A4:** Variance explained by the four EFA models

| Indices | Raw indices $x$ | $\ln(x)$ | $\ln(x+1)$ | $\sqrt{x}$ |
|---|---|---|---|---|
| $h$ | 0.981 | 0.926 | 0.932 | 0.961 |
| $m$ | 0.921 | 0.795 | 0.804 | 0.866 |
| $g$ | 0.998 | 0.993 | 0.993 | 0.996 |
| $h(2)$ | 0.949 | 0.975 | 0.975 | 0.960 |
| $A$ | 0.999 | 0.999 | 0.999 | 0.999 |
| $R$ | 0.999 | 0.995 | 0.995 | 0.997 |
| $h_w$ | 0.999 | 0.999 | 0.999 | 0.999 |

**Table A5:** Variance explained by the four EFA models as in Table A4, but comprising $N$ and $C$

| Indices | Raw indices $x$ | $\ln(x)$ | $\ln(x+1)$ | $\sqrt{x}$ |
|---|---|---|---|---|
| $h$ | 0.971 | 0.948 | 0.951 | 0.957 |
| $m$ | 0.916 | 0.783 | 0.790 | 0.859 |
| $g$ | 0.998 | 0.995 | 0.995 | 0.997 |
| $h(2)$ | 0.939 | 0.931 | 0.935 | 0.938 |
| $A$ | 0.953 | 0.931 | 0.931 | 0.941 |
| $R$ | 0.999 | 0.996 | 0.997 | 0.999 |
| $h_w$ | 0.999 | 0.999 | 0.999 | 0.999 |
| $N$ | 0.835 | 0.957 | 0.956 | 0.913 |
| $C$ | 0.999 | 0.999 | 0.999 | 0.999 |

**Table A6:** Confirmatory factor analysis matrix for the raw indices, but comprising $N$ and $C$

|        | Raw indices $x$ | | $\ln(x)$ | | $\sqrt{x}$ | |
| Indices | Component | | Component | | Component | |
|        | 1 | 2 | 1 | 2 | 1 | 2 |
|--------|------|------|------|------|------|------|
| $h$    | 0.97 |      | 0.98 |      | 1.00 |      |
| $m$    | 0.95 |      |      | 0.87 | 0.60 | 0.81 |
| $g$    | 1.00 |      |      | 1.00 | 0.52 | 0.52 |
| $h(2)$ | 0.96 |      | 0.96 |      | 0.95 |      |
| $A$    |      | 1.03 | 0.95 |      |      | 1.00 |
| $R$    | 1.00 |      |      | 1.00 | 0.51 | 0.53 |
| $h_w$  | 1.00 |      |      | 1.00 | 0.51 | 0.53 |
| $N$    | 0.74 |      | 0.84 |      | 0.81 |      |
| $C$    |      | 0.83 |      | 0.81 |      | 0.85 |

**Table A7:** Summary statistics of the CFA model fit

| Indices | Raw indices $x$ | | $x, \ln(x), \sqrt{x}$ | $x$ | $\ln(x)$ | $\sqrt{x}$ |
|         | Unstandardized loadings | Standard error | p-value | $R^2$ | $R^2$ | $R^2$ |
|---------|------|-------|--------|------|------|------|
| $h$     | 6.69 | 1.01  | <0.05  | 0.94 | 0.97 | 0.99 |
| $m$     | 6.46 | 1.90  | <0.05  | 0.91 | 0.75 | 0.84 |
| $g$     | 7.06 | 1.70  | <0.05  | 0.99 | 0.99 | 0.99 |
| $h(2)$  | 6.58 | 0.23  | <0.05  | 0.93 | 0.93 | 0.93 |
| $A$     | 7.52 | 2.45  | <0.05  | 0.99 | 0.92 | 0.99 |
| $R$     | 7.07 | 1.53  | <0.05  | 0.99 | 0.99 | 0.99 |
| $h_w$   | 7.06 | 1.30  | <0.05  | 0.99 | 0.99 | 0.99 |
| $N$     | 4.33 | 13.52 | n.s.   | 0.55 | 0.70 | 0.66 |
| $C$     | 5.10 | 0.74  | <0.05  | 0.69 | 0.65 | 0.75 |

**Appendix B: Definition of the discussed indices**

The h-index is the highest number $h$ of articles that each received $h$ or more citations (*Hirsch, 2005*).

The h(2)-index (*Kosmulski, 2007*) is the highest number $h(2)$ of articles that each received $[h(2)]^2$ or more citations.

The g-index is the highest number $g$ of articles that together received $g^2$ or more citations (*Egghe, 2006b*). This is equivalent to the highest number of articles that received $g$ or more citations on average (*Schreiber, 2010a*).

The m-index (*Bornmann et al., 2008a*) is defined as the median number of citations received by the articles included in the h-core.

The A-index (*Jin, 2006*) is defined as the average number of citations received by the articles included in the h-core.

The R-index (*Jin et al., 2007*) is given by the square root of the total number of citations received by the articles included in the h-core. This is equivalent to $R = \sqrt{Ah}$.

The $h_w$-index (*Egghe and Rousseau, 2008*) is given by the square root of the total number $s$ of citations received by the highest number of articles that each received $s/h$ or more citations.